\documentclass[openacc]{rstransa}

\usepackage{algorithm}
\usepackage{algorithmic}

\newcommand{\xx}{\mathbf{x}}
\newcommand{\yy}{\mathbf{y}}
\newcommand{\zz}{\mathbf{z}}

\newcommand{\MM}{\mathcal{M}}
\newcommand{\beps}{\boldsymbol{\epsilon}}
\newcommand{\btheta}{\boldsymbol{\theta}}
\newcommand{\bsigma}{\boldsymbol{\sigma}}

\begin{document}

\title{Combining data assimilation and machine learning to infer unresolved scale parametrisation.
}

\author{
Julien Brajard$^{1,2}$, Alberto Carrassi$^{3,4}$, Marc Bocquet$^{5}$ and Laurent Bertino$^{1}$}

\address{$^{1}${Nansen Center (NERSC), 5006, Bergen, Norway}\\
$^{2}${Sorbonne University, Paris, France}\\
$^{3}${Department of Meteorology, University of Reading and NCEO, United-Kingdom}\\
$^{4}${Mathematical Institute, University of Utrecht, The Netherlands}\\
$^{5}${CEREA, joint laboratory \'Ecole des Ponts ParisTech and EDF R\&D, 
Universit\'e Paris-Est, France}}

\subject{Computer modelling and simulation; artificial intelligence; climatology.}


\corres{Julien Brajard\\
\email{julien.brajard@nersc.no}}

\begin{abstract}
In recent years, machine learning (ML) has been proposed to devise data-driven parametrisations of unresolved processes in dynamical numerical models. In most cases, the ML training leverages high-resolution simulations to provide a dense, noiseless target state. Our goal is to go beyond the use of high-resolution simulations and train ML-based parametrisation using direct data, in the realistic scenario of noisy and sparse observations.

The algorithm proposed in this work is a two-step process. First, data assimilation (DA) techniques are applied to estimate the full state of the system from a truncated model. The unresolved part of the truncated model is viewed as a model error in the DA system. In a second step, ML is used to emulate the unresolved part, a predictor of model error given the state of the system. Finally, the ML-based parametrisation model is added to the physical core truncated model to produce a hybrid model.

The DA component of the proposed method relies on an ensemble Kalman filter while the ML parametrisation is represented by a neural network.
The approach is applied to the two-scale Lorenz model and to MAOOAM, a reduced-order coupled ocean-atmosphere model.   
We show that in both cases the hybrid model yields forecasts with better skill than the truncated model. Moreover, the attractor of the system is significantly better represented by the hybrid model than by the truncated model.
\end{abstract}










\maketitle

\section{Introduction}

The Earth climate system is one example of a natural system that is reasonably well represented through known physical laws and that has been intensively observed for decades (see, {\it e.g.},~\cite{merchant2019satellite}). Physical laws, in the form of ordinary (ODEs) or partial differential equations (PDEs), are implemented through numerical models providing the time evolution of the system's state. Although weather and climate predictions have constantly improved, and will likely continue to do so, uncertainties will ineluctably remain. Those usually fall into two major classes: (i) the internal variability driven by the sensitivity to the initial conditions, and (ii) the model errors. The former has to do with the amplification of the initial condition error and arises even in perfect models - it is mitigated by using data assimilation (DA)~\cite{carrassi2018data}. The latter is present even if one would perfectly know the initial conditions and has to do with the incorrect and/or incomplete representation of the laws governing the system. The two sources of errors are inevitably entangled and it is difficult to separate them in practice.

Machine learning (ML) was recently shown to be effective in reducing model error, in particular that originating from unresolved scales. This has been achieved using two approaches. The first consists in learning a subgrid parameterisation of a model from existing physics-based expensive parametrisation schemes~\cite{o2018using, rasp2018deep}, or from the differences between high- and low-resolution simulations~\cite{krasnopolsky2013using, bolton2019applications, brenowitz2018prognostic, rasp2020coupled}. In those approaches, the unresolved part of the model is represented by a ML process while the core of the model is derived from ODEs.
The second approach is to emulate the entire model using observations. With spatially dense and noise-free data, this approach has been based on sparse regression~\cite{brunton2016discovering}, echo state networks~\cite{pathak2018model, faranda:hal-02337839}, recurrent neural networks~\cite{park2010time}, residual neural network~\cite{fablet2018bilinear} or convolutional neural networks~\cite{scher2018toward, dueben2018challenges}. The challenging problem of partial and/or noisy observations has been addressed using dedicated NN architecture~\cite{de2019deep} or in combination with data assimilation methods
~\cite{nguyen2019like,laloyaux2020towards, bonavita2020, brajard2020combining, bocquet2020bayesian}. 

This work presents a new method to obtain a data-driven parameterisation of a model's unresolved scale. In particular, we aim at producing a hybrid model combining the physics-based core (encoding the best of our knowledge of the resolved scales physics) with the data-driven parameterisation. By leveraging the use of DA, our method efficiently handles noisy and sparse observations.   


\section{\label{sec:statements}Objectives and definitions}

\subsection{Statement of the problem}
We consider an autonomous chaotic dynamical system, seen as our reference ``truth'', represented by the ODE
\begin{align}
\label{eq:ode}
\frac{\textrm{d}\mathbf{z}(t)}{\textrm{d}t} = f(\mathbf{z}(t)),
\end{align}
with $\zz(t) \in \mathbb{R}^{N_z}$ being the system's state at time $t$.
From an arbitrary state $\zz(t)$ on the system's attractor the model can be integrated forward for one time step $\delta t$, to get:
\begin{align}
\label{eq:true-model}
\zz_{\delta t} = \MM(\zz, \delta t),
\end{align}
where $\zz_{\delta t}=\zz(t+\delta t)$.

Let us formally define a projection operator $\Pi:  \mathbb{R}^{N_z} \mapsto \mathbb{R}^{N_x}$ such as $\xx = \Pi(\zz)$, with $\xx$ being the projection of the full state into a reduced dimension state: $N_x<N_z$. For example, $\Pi$ can be a subsampling operator retaining only a subset of $\zz$ or a downscaling operator from a high-resolution state to a lower resolution. From Eq.\,\ref{eq:true-model}, we also define $\xx_{\delta t}=\Pi(\zz_{\delta t})$.

Consider now a scale-truncated model, $\MM^{\rm r}$, that provides an imperfect description of Eq.\,\eqref{eq:true-model} in the reduced space $\mathbb{R}^{N_x}$,
\begin{align}
\label{eq:trunc-model}
\xx^{\rm r}_{\delta t} = \MM^{\rm r}(\xx^{\rm r}, \delta t),
\end{align}
where the superscript ${\rm r}$ stands for ``resolved''. 
When initialised from the projection of the truth into the reduced space ({\it i.e.} no initial condition error: $\xx^{\rm r} =\xx$), the difference between the 1-time step predictions from Eqs.~\eqref{eq:true-model} and \eqref{eq:trunc-model} defines the model error due to the neglected scales in the resolved model
\begin{align}
\begin{split}
\beps^{\rm m}(\zz,\delta t) = \xx_{\delta t} - \xx^{\rm r}_{\delta t} = \Pi\circ\MM(\zz, \delta t) - \MM^{\rm r}(\Pi(\zz), \delta t).\\
\end{split}
\end{align}
The objective of this work is to complement the resolved model \eqref{eq:trunc-model} by an empirical representation of the unresolved scales
based on a neural network (NN) trained on incomplete and noisy data. We will hereafter denote the NN-based unresolved scale representation by $g(\xx,\btheta)$. The vector $\btheta$ is the set of trainable parameters of the NN and is determined by minimising the loss function
\begin{align}
\label{eq:cont-loss}
L(\btheta) = \mathbb{E}_{\zz \sim p_Z} \left[g(\Pi(\zz),\btheta) - \beps^{\rm m}(\zz,\delta t)\right]^2,
\end{align}
where $p_Z$ is the invariant probability density function (assumed to exist) on the attractor.

We construct the hybrid model $\MM^{\rm h}$, parametrised by $\btheta$, such that
\begin{align}
\begin{split}
\xx^{\rm h}_{\delta t} = \MM^{\rm r}(\xx, \delta t) + g(\xx,\btheta)
=\MM^{\rm h}(\xx, \delta t).
\end{split}
\end{align}
Our objective is to optimally estimate the parameters $\btheta$ so that the hybrid model is the most accurate representation of the true underlying dynamics.

Apart from trivial cases, the loss function in Eq.\,\eqref{eq:cont-loss} cannot be computed: $\xx_{\delta t}$ is unknown and $p_Z$ is generally intractable. Assuming ergodicity of the true dynamics, we can however estimate it by a Monte-Carlo approach such that:
\begin{align}
\label{eq:mc-loss}
L(\btheta) \approx \frac{1}{K}\sum_{k=0}^{K-1} \left[g(\xx_k,\btheta) - \beps^{\rm m}(\zz_k,\delta t)\right]^2,
\end{align}
where $\zz_{0:K-1} = \{\zz_0, \zz_1, \cdots, \zz_{K-1}\}$, $\zz_k=\zz(t_k)$, is a set of samples of $p_Z$, typically extracted from a time series of modelled state variables and $\xx_k = \Pi(\zz_k)$.
Such samples are usually not independent (due to the underlying dynamics being deterministic), and only provide a biased approximation of $p_Z$. Furthermore, the need to sample the whole attractor implies treating time series significantly longer than the decorrelation time ({\it i.e.} $K$ very large in general) .


\subsection{\label{statement:framework}Framework of the study}
The loss function, Eq.\,\eqref{eq:mc-loss}, cannot be minimised directly because some of its key entries are unavailable, in particular, obviously, the true process Eq.\,\eqref{eq:true-model} and the time series $\zz_{0:K-1}$. The available terms in the loss function are:\\

\noindent {\bf The truncated model.}
The truncated model, our best available knowledge about the true process, is usually very complex (high dimensional, nonlinear and with diagnostic variables). Hence, we shall assume that its gradient, $\nabla_{\xx}{\MM^{\rm r}}$, cannot be computed analytically. 
We will thus focus on developing a (model) adjoint-free approach that is more flexible and suitable to high dimensional nonlinear scenarios where deriving and maintaining an adjoint model is a difficult and costly task \cite{tian2018nonlinear}. When the gradient can be computed, other efficient methods exist~\cite{bocquet2020bayesian}.\\

\noindent {\bf Observations.}
Observations are incomplete and noisy and are obtained through:
\begin{align}
\label{eq:obs}
\yy_k = \mathbf{H}_k\xx_k + \beps^{\rm o}_k,
\end{align}
where $\xx_k$ is the true state in the reduced space, $\yy_k \in \mathbb{R}^{N_y} $ and $\mathbf{H}_k \in \mathbb{R}^{N_y\times N_x}$ are the observation vector and operator respectively at $t_k$, while $\beps^{\rm o}_k$ is the observation error, assumed to be uncorrelated in time and normally distributed with mean $0$ and a variance-covariance matrix $\left(\sigma^{\rm o} \right)^2\mathcal{I}_{N_y}$,  where $\mathcal{I}_{N_y}$ is the identity matrix of size $N_y \times N_y$ and $\sigma^{\rm o}$ is the standard deviation. For simplicity, the observation error standard deviation is taken constant and the observation operator linear. Both assumptions can be relaxed without major drawbacks even if it can induce practical difficulties. The ideal, most favourable, situation in which the full system's state is observed in the reduced space with no error
is referred as the "perfect observation" case: $\yy_k = \xx_k$. For convenience, we further assume that observations are available regularly at multiples of the model time step such that $\Delta t = t_{k+1}-t_k = N_c \delta t$, $N_c \in \mathbb{N}^*$. This also accounts for the fact that, in general, the observation sampling period is longer than the integration time step of the numerical model.

\section{\label{sec:method}Method}

\subsection{\label{method:approx}Loss function approximation}
Let us assume that an estimation of $\xx_k$~($k \in \{0,\cdots,K\}$) is available at observation times, every $\Delta t = N_c \delta t$, so that
\begin{align}
\xx_{k+1} = \Pi\circ\mathcal{M}^{(N_c)}(\zz_k,\delta t),
\end{align}
where $\mathcal{M}^{(N_c)}$ is the composition of the model $N_c$ times, and similarly for the truncated model, $\xx^{\rm r}_{k+1}=\mathcal{M}^{\rm{r}(N_c)}(\xx_k,\delta t)$. Since observations are not available at each time step ($N_c>1$), the model error $\beps^{\rm m}$ is not known at each time step neither, and the loss function Eq.\,\eqref{eq:mc-loss} cannot be exactly computed.
In the following, we will present two key simplifying assumptions that will lead to a tractable approximation of the loss function.

The first consists in assuming $\Delta t$ to be short enough so that the state evolved by the truncated model is independent of the model error due to the unresolved scale so that the model error is an additive term to the truncated model forecast after a $\Delta t$ time integration. The second in that the variability of $\beps^{\rm m}$ is small within $\Delta t$. The combination of these two hypotheses is known as the {\it linear superposition assumption}, and can be formalised as:
\begin{align}
\xx_{k+1} \approx \xx^{\rm r}_{k+1} + N_c\times \beps^{\rm m}(\zz_k,\delta t).
\end{align}
Note that the same approximation was made in a similar setting by~\cite{rasp2020coupled}.

The optimisation can now be performed using the approximate loss function:
\begin{equation}
\label{eq:actual-loss}
L(\btheta) \approx L^{\rm a}(\btheta) = \frac{1}{K N_c^2}\sum_{k=0}^{K-1} \left[N_c g(\xx_k,\btheta) - (\xx_{k+1} - \xx^{\rm r}_{k+1})\right]^2.
\end{equation}
The modified loss function, Eq.\,\eqref{eq:actual-loss} can be computed without knowing the full state $\zz_k$ but only its projection $\xx_k$. $L^{\rm a}$ can be minimised using a gradient descent algorithm as long as the gradient of $g(\xx_k,\btheta)$ can be computed, which is standard for any neural network library.

\subsection{\label{method:algo}Description of the algorithm}
In order to minimise the loss function defined in Eq.\,\eqref{eq:actual-loss}, a sequence of $\xx_{0:K}$ has to be available. Two cases are considered: the first is the aforementioned "perfect observations" case, in which we have a complete and noise-free sequence of a state variable $\xx_k$ in the reduced space. This ideal situation will set the upper-bound performance in the algorithm evaluation that follows. The second case is the more realistic case of noisy (yet unbiased) and possibly incomplete observations. Here, a complete sequence $\xx_{0:K}$ is obtained by processing the incomplete and noisy observations using DA~\cite{carrassi2018data}: the observations $\yy_k$ are combined with the forecast from the truncated model $\xx^{\rm r}_k$ in order to provide the analysed state vector $\xx^{\rm a}_k$. The DA method used in this work is the Finite-Size Ensemble Kalman Filter (EnKF-N)
~\cite{bocquet2015expanding} implemented in the DAPPER framework~\cite{raanes2018dapper}. Even if the proposed algorithm is general and suitable for any DA algorithm, the EnKF-N has been chosen because of its efficiency. In particular, the inflation factor, a needed fix to mitigate the impact of sampling errors in the ensemble-based DA methods, is automatically estimated (thus avoiding long tuning). This inflation factor accounts also implicitly and partially for the effect of the model error.

The correction $\xx^{\rm a}_k - \xx^{\rm r}_k$ made by DA is called analysis increment and was used to estimate model error due to unresolved scales in sequential DA in \cite{carrassi2011treatment,mitchell2015accounting}. An analysis increment is composed of a correction both for the model error (which is what the hybrid model aims at estimating) and for the initial conditions error (which cannot be represented by the hybrid model). There is also an additional uncertainty due to the observation errors. The initial conditions and the observations errors are two sources of uncertainties called the data uncertainties. If they are too large, relatively to the model error, they could bias the estimation of the loss function given in Eq.~\eqref{eq:actual-loss} causing the hybrid model to overfit on these data and to lack generalisation to other initial conditions. To mitigate this problem, the time series $\xx^{\rm a}_{0:K}$ estimated by DA is filtered using a simple low-pass filter (a rolling mean) producing a smoothed time series $\xx^{\rm s}_{0:K}$. This filter is expected to correct for data uncertainty provided that the observation error is uncorrelated in time, and thus contains high-temporal frequencies. On the other hand, this filtering could remove the fastest scales of the unresolved part of the model. Although it has been assumed in the linear superposition assumption that these high-frequencies can be neglected, we do not know a priori to which extent, which can lead to possibly hamper the forecast skill. The scale separation between the model error acting on long time scales and the initial errors acting on faster scales has been used in previous studies either to estimate the model error in DA~\cite{laloyaux2020towards} or to improve the forecast of a NN model~\cite{faranda:hal-02337839}.
Finally, note that the filtering can be adapted separately for the fast and the slow variables contained in $\xx_k$.
This is, for instance, the case in coupled atmosphere-ocean models, and is addressed in section~\ref{sec:maooam}.
\begin{algorithm}
\caption{Summary of the algorithm used to determine the hybrid model}
\label{algo:main}
\begin{algorithmic}[1]
\REQUIRE Observations $\yy_{0:K}$,\\ truncated model $\MM^{\rm r}$,\\ NN architecture $g(\xx,\btheta)$.
\ENSURE state vector estimation $\xx^{\rm s}_{0:K}$, \\
optimal value of $\btheta$.
\IF {$\yy_{0:K}$ is perfect}
\STATE $\xx^{\rm s}_{0:K} = \yy_{0:K}$
\ELSE
\STATE Use a DA algorithm (e.g. EnKF-N) to estimate the state vector series
$\xx^{\rm a}_{0:K}$
\STATE Filter the components (or a subset of components) of $\xx^{\rm a}_{0:K}$ using a low-pass filter to produced the smoothed field $\xx^{\rm s}_{0:K}$
\ENDIF
\STATE compute the target for the NN: $\beps^{\rm m}_k =(1/N_c)(\xx^{\rm s}_{k+1} -\MM^{\rm r}(\xx^{\rm s}_k, N_c\delta t))$
\STATE Determine $\btheta$ (training of the NN) using the dataset $(\xx^s_{0:K-1};\beps^{\rm m}_{0:K-1})$
\RETURN $(\xx^{\rm s}_{0:K},\btheta)$
\end{algorithmic}
\end{algorithm}

In both perfect and imperfect observations cases, the loss function $L^{\rm a}(\btheta)$ is minimised using a standard NN training procedure: if $g(\xx_k,\btheta)$ is represented by a NN, $\xx_k$ as the inputs and $\btheta$ as weight, the problem is equivalent to a supervised regression problem in which the targeted output of the neural network is $\beps^{\rm m}_k =(1/N_c)(\xx_{k+1} - \xx^{\rm r}_k)$. The algorithm is summarised in Algorithm~\ref{algo:main}.

\subsection{Numerical experiment protocol}
The performance of the algorithm is evaluated on twin experiments: a full model $\MM$ is used to produce true states, synthetic observations, and to assess the forecast skill of the hybrid model. The truncated model $\MM^{\rm r}$ is obtained by neglecting some components of the true model.

A series of true states $\xx_{0:K}$ is produced using the true model after projection in the reduced space. This series is used as perfect observations to obtain the so-called "perfect observation-derived hybrid model". Then, synthetic observations are generated using the definition in Eq.~\eqref{eq:obs}. Here, it is assumed that the observation operator and the observation error statistics are perfectly known. This assumption is not needed for the functioning of the proposed algorithm and could be relaxed at a future stage. Nonetheless, the assumption is done here to simplify the interpretation of the results so the focus is on the truncated model error. Furthermore, the observation error statistics can also be estimated within the DA process itself (see \cite{Tandeo2020} for a review of such methods), and thus be integrated into our combined ML-DA method. 

The observations generated using Eq.~\eqref{eq:obs} are used in Algorithm~\ref{algo:main} to produce the so-called "DA-derived hybrid model". Note that the perfect observation-derived hybrid model benefits from the optimal information at a given time step (the state is completely observed without noise). Given the conditions stated in section~\ref{sec:statements}\ref{statement:framework}, it is expected to be the best possible model and represents the benchmark against which we will test the DA-derived hybrid model.

\subsection{\label{method:metrics}Evaluation metrics}
The skill of each model is evaluated by running an ensemble of $N^{\rm f}=20$ forecasts for  $N^{\rm f}$ different initial conditions and for a time period $\tau$: $\xx^{{\rm f}(l)}(\tau)$ ($l=1,\cdots,N^{\rm f}$). The $N^{\rm f}$ different initial conditions are chosen on the attractor of the true model (by running the model long enough before setting the initial conditions), independent of the values of the time series $\xx_{0:K}$ used to perform the DA and the NN training. This ensemble of forecast constitutes the so-called test dataset, as it is not used to optimise nor to tune the algorithm. If tuning the algorithm is needed, the original time series $\xx_{0:K}$ can be split in a training part (used to optimise the NN) and a validation part (used to tune the NN algorithm).

As evaluation metric we will use the relative root mean square error (R-RMSE)
\begin{align}
\label{eq:r-rmse}
\operatorname{R-RMSE}(\tau,n) = \sqrt{\frac{1}{N^{\rm f}} \sum_{l=1}^{N^{\rm f}} \frac{1}{2 V_n(\xx^{\rm t})}\left( x^{{\rm f}(l)}_n(\tau) - x^{{\rm t}(l)}_n(\tau) \right)^2},
\end{align}
where $x^{{\rm f}(l)}_n(\tau)$ (resp. $x^{{\rm t}(l)}_n(\tau)$) is the forecast from the hybrid or the truncated model (resp. the true model) at time $\tau$ for the $n$-th component of $\xx$ and for the sample $l$ corresponding to a simulation for one particular initial condition. $V_n(\xx^{\rm t})$ is the $n$-th component of the variance over the time dimension of the true model forecast time series $\xx^{\rm t}$.
Note that if the truncated and/or the hybrid models have the same variability as the true one (i.e. the same variance in time), the R-RMSE converges to 1. 

\section{\label{sec:luv}Application to the two-scale Lorenz model}

\subsection{Description of the model}
The two-scale Lorenz model~\cite{lorenz2005designing}, hereafter L2S, is given by the following set of ODEs:
\begin{align}
\begin{split}
\frac{{\rm d} x_n}{{\rm d}t} =&
\psi^+_n(\xx) + F - \frac{c}{b}\sum_{m=0}^9u_{m+10n}\\
\frac{{\rm d} u_m}{{\rm d}t} =& \frac{c}{b}\psi^-_m(b{\bf u}) + h\frac{c}{b}x_{m/10},\\
\psi_n^\pm(\xx) = & x_{n\mp 1}(x_{n \pm 1} - x_{n \mp 2}) - u_n,
\end{split}
\end{align}
where $n=0,\cdots,N_x-1$ ($N_x=36$) and $m=0,\cdots, N_u-1$ ($N_u=360$). The indices $n$ are periodic, e.g., $x_0=x_{N_x}$. The values chosen for the parameters are the same as in~\cite{bocquet2019data}: the time scale ratio $c$ is set to 10, the space-scale ratio $b=10$, the coupling $h=1$ and the forcing $F=10$. Time $t$ is expressed in model time unit, denoted MTU hereafter.

This set of two-scale ODEs, considered as the true model, is integrated using a fourth-order Runge-Kutta scheme with a time step of 0.005 MTU. The ODEs describing the evolution of $\xx$ only represent the truncated model and are obtained by setting the coupling $c$ to 0. It is integrated using a fourth-order Runge-Kutta scheme with a time step of 0.01 MTU.

\subsection{\label{luv:setting} Setup of the reference experiment}
A so-called "reference experiment" is defined in this section.
The true model is integrated over approximately 1500 MTU after a spinup of 3 MTU to produce the true state, on which observational noise is added.
The EnKF-N is used to assimilate these observations using a large number $N=50>N_x$ of ensemble members to reduce sampling errors. A noise is added to the state vector after each forecast to approximately account for the model error due to the model being truncated. It helps to avoid filter divergence and can be seen as additive inflation. This step is necessary given that, due to model error, the forecast ensemble would be otherwise under-dispersive.  The noise is assumed Gaussian with zero mean and standard deviation $\bsigma^{\rm m} = {\bf 0.06}$ optimised after tuning experiments (not shown here).
In this reference experiment, the analysis obtained from the DA is not filtered, yet the sensitivity to the filtering of the DA analysis is studied in section~\ref{sec:luv}\ref{luv:sensitivity}.

The last step of the algorithm is to train a neural network to emulate the unresolved part of the model on the 1500 MTU time series produced by DA.
The NN architecture is composed of convolutional layers (denoted ``conv.'' in Table~\ref{tab:exp-setting}) where the non-linear activation function is a hyperbolic tangent (denoted "tanh").
Some additional parameters have been added, mainly to regularise the training: a batchnorm layer at the input layer, which normalises the training batch, and a L2-regularisation term on the parameters of the last layer. The parameters of the NN are optimised using the "RMSprop"~\cite{hinton2012neural} optimiser over 100 epochs. For each epoch, batches of 33 training examples are used to optimise the weights, until all the examples are consumed: this is a standard stochastic minimisation procedure~\cite{bottou2010large}.
Full details on the reference experiment are given in Table~\ref{tab:exp-setting}, in the column labelled L2S.

\begin{table}[!ht]
\caption{Settings of the numerical experiments with the L2S ``reference experiment'' and with MAOOAM.}
\label{tab:exp-setting}
\begin{tabular}{cc|c|cc}
\hline
&&L2S&\multicolumn{2}{c}{MAOOAM}\\
Parameter & Symbol & Value & Value & Note \\
\hline
climatological std & $\bsigma^{\rm hf}$ & NA & & calculation in~\ref{sec:maooam}\ref{maooam:setting}.\\
\hline
\multicolumn{5}{c}{\bf Model parameter}\\
Size of the state & $N_x$ &36 &36 &\\
Integration time step & $\delta t$ & 0.01 MTU &1.6 min & \\
Integration time & $T$ & 1500 MTU & 62 years. \\
\hline
\multicolumn{5}{c}{\bf Imperfect observation setting}\\

Standard deviation & $\bsigma^{\rm o}$ & 0.1 & $0.1 \bsigma^{\rm hf}$ & Ocean filtered.\\
Observation operator & $H$ &$\mathcal{I}_{N_x}$ &$\mathcal{I}_{N_x}$ &Identity matrix\\
Time sampling & $\Delta t$ & 0.05 MTU & 27 hours &\\
\hline
\multicolumn{5}{c}{\bf Data assimilation}\\

DA algorithm & EnKF-N && \\
Ensemble size & $N$ & 50& 50 &\\
Model additive noise & $\bsigma^{\rm m}$& 0.06
& $10^{-3} \bsigma^{\rm hf}_{1:20}$ & Atmosphere only \\
Low-pass filtering size & &No& 55 days & Ocean only. \\
\hline
\multicolumn{5}{c}{\bf Neural Network}\\
&\multicolumn{2}{|c|}{L2S} & MAOOAM & \\
Type of Layer 1 & \multicolumn{2}{|c|}{Batchnorm} & Batchnorm & \cite{ioffe2017batch}\\
Type of Layer 2 & \multicolumn{2}{|c|}{conv.}& dense & \\
Size of Layer 2 & \multicolumn{2}{|c|}{43 } & 100 &\\
Activation of Layer 2 & \multicolumn{2}{|c|}{tanh} & ReLU & \cite{glorot2011deep} \\
Filter size of Layer 2& \multicolumn{2}{|c|}{5} & - \\
Type of Layer 3 & \multicolumn{2}{|c|}{conv.} & dense & \\
Size of Layer 3 & \multicolumn{2}{|c|}{28} & 50 & \\
Filter size of Layer 2& \multicolumn{2}{|c|}{1} & - \\

Activation of Layer 3 & \multicolumn{2}{|c|}{tanh} & ReLU & \\
Type of Layer 4 & \multicolumn{2}{|c|}{conv.} & dense & \\
Size of Layer 4 & \multicolumn{2}{|c|}{36 } & 36& \\
Activation of Layer 4 & \multicolumn{2}{|c|}{Linear} & Linear & \\
L2 regularisation & \multicolumn{2}{|c|}{0.07} & $10^{-4}$& \\
Optimiser & \multicolumn{2}{|c|}{RMSprop} &RMSprop & \cite{hinton2012neural}\\
Number of epochs & \multicolumn{2}{|c|}{100} & 100& \\
batch size & \multicolumn{2}{|c|}{33}& 128& \\

\hline
\end{tabular}
\vspace*{-4pt}
\end{table}

\subsection{Results}
\begin{figure}[!ht]
\centering
\includegraphics[width=.6\textwidth]{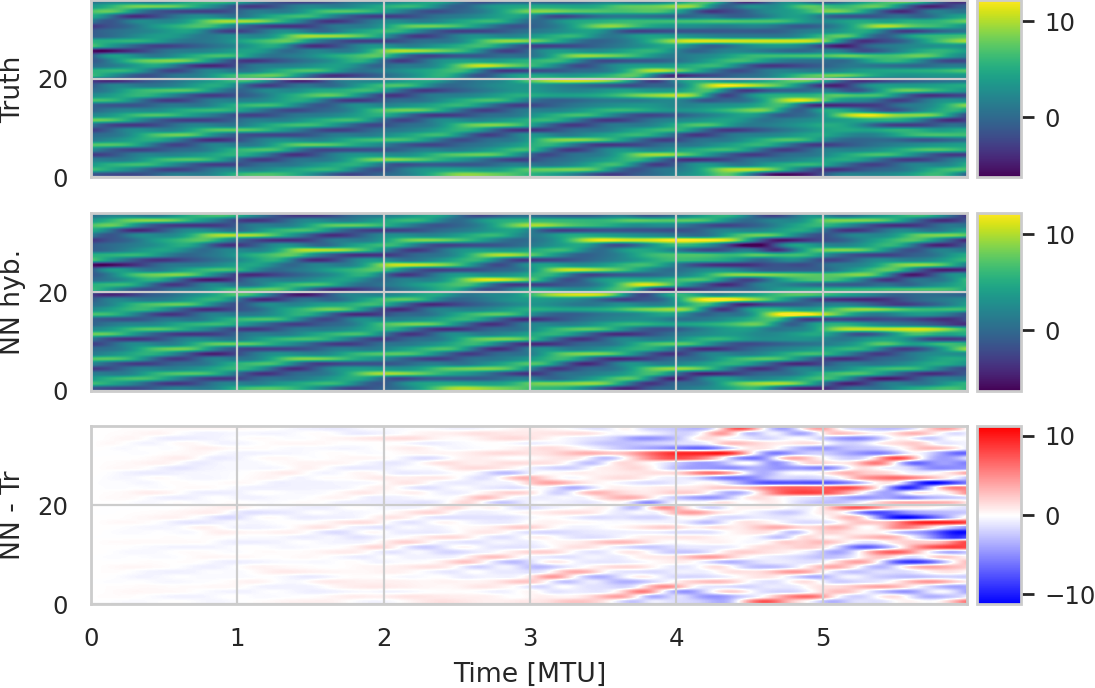}

\caption{Hovm\o{}ller plot of the true model (upper panel) and of the DA-derived hybrid model (middle panel) for the same particular initial conditions over 5 MTU. The bottom panel shows the difference between the two simulations. The setup of the experiment is detailed in Table~\ref{tab:exp-setting}.}
\label{fig:luv-sim}
\end{figure}

The terminology of the experiments described in the following is recalled in table~\ref{tab:luv-terminology}.
In Figure~\ref{fig:luv-sim}, both the true model and the DA-derived hybrid model (based on the reference experiment described in section~\ref{sec:luv}\ref{luv:setting}) are initialised from an initial condition on the attractor, chosen to be independent of the training set $\xx_{0:K}$. The true and the hybrid model are run over 5 MTU, and their difference is displayed. It can be noticed that both runs are very close until 2 MTU and that the hybrid model has predictive skill until 3-4 MTU for this particular set of initial conditions. Note that the Lyapunov time of the truncated model is 0.72
~\cite{carlu2018lyapunov}, meaning that the hybrid model provides accurate forecasts until 3 Lyapunov times.

\begin{table}[!ht]
\caption{Numerical experiments terminology}
\label{tab:luv-terminology}
\begin{tabular}{p{.3\textwidth}p{.6\textwidth}}
\hline
Reference experiment & Setup defined in table~\ref{tab:exp-setting} (column L2S). \\
\hline

DA-derived hybrid model & NN trained with data assimilation reanalysis obtained from noisy and sparse observations.\\
\hline
Perfect-observation-derived hybrid model & NN trained with perfect observations.\\
\hline

\end{tabular}
\vspace*{-4pt}
\end{table}

\begin{figure}[!ht]
\centering
\includegraphics[width=\textwidth]{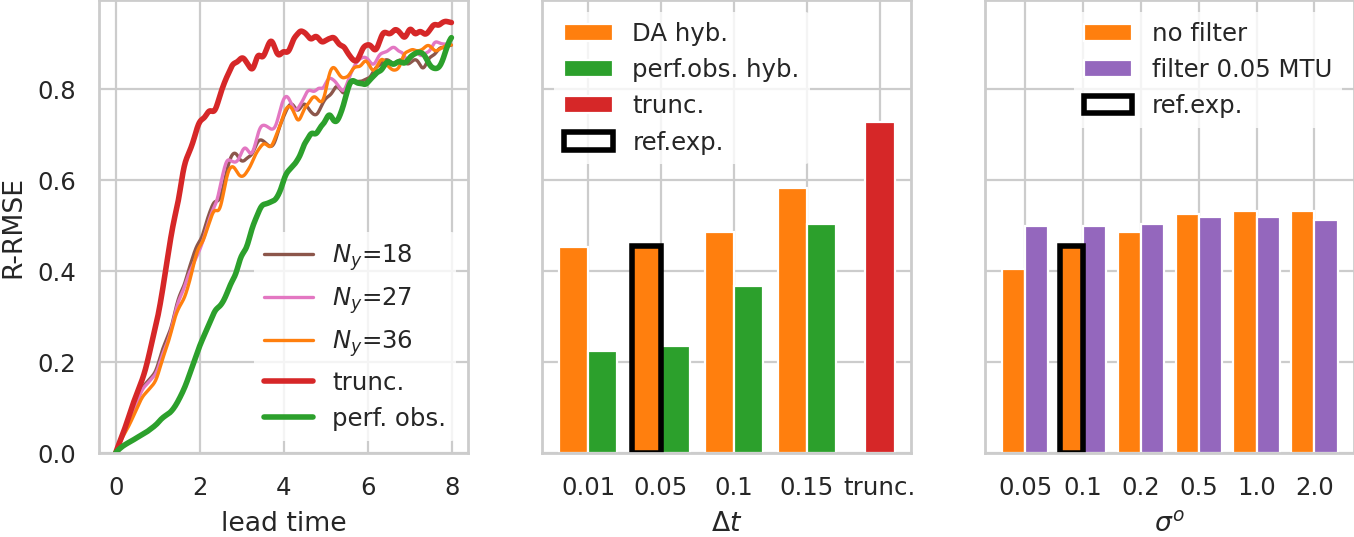}
\caption{{\bf Left}: R-RMSE versus time for the perfect observations-derived hybrid model (green), the truncated model (red) and the DA-derived hybrid model (other colours) for 3 different densities of observation. Observation standard deviation, $\sigma^o=0.1$, the time interval, $\Delta t=0.05~MTUs$, are as in the reference experiment; thus for $N_y=36$ we retrieve the reference experiment. {\bf Middle}: R-RMSE at a lead time $\tau=2$~MTUs for the DA-derived hybrid models (orange) and the perfect observation-derived hybrid models (green) for different observation sampling time $\Delta t$ as well as the truncated model (red). {\bf Right}: R-RMSE for the hybrid models trained with no filtering of DA analysis (orange) and with a 0.05~MTU window filter (purple) for different observation error standard deviation $\sigma^{\rm o}$. Black contour indicates the reference experiment conditions described in Table~\ref{tab:exp-setting}.}
\label{fig:luv-p}
\end{figure}

In Figure~\ref{fig:luv-p}, the R-RMSE is averaged over 20 members corresponding to  20 initial conditions and also across all the 36 components of $\xx$. R-RMSE is displayed as a function of time. 
Several densities of observations have been considered: if $N_y=36$ the full state is observed in the reduced space at each observation time. If $N_y<36$, $\mathbf{H}_k$ is a sub-sampling operator that draws randomly $N_y$ values from the state following a uniform distribution changing the observation locations at each time step.

Results shown in Figure~\ref{fig:luv-p} (left panel) confirms that the DA-derived hybrid model has a predictive skill, significantly better than the truncated model until 4 MTU. The effect of reduced observation density is minor: the skill of the various hybrid models' forecasts is very similar. This shows the algorithm efficiency in handling sparse data to accurately train a NN model. This is a key strength of our method; most of the other approaches that parametrise a part of the model using ML assume dense observations, e.g.~\cite{rasp2018deep, brenowitz2018prognostic,o2018using} (similarly to our perfect observation case). 

\subsection{\label{luv:sensitivity}Sensitivity studies}
In Figure~\ref{fig:luv-p} (middle and right panels), the forecast sensitivity to different parameters is studied using the R-RMSE at a lead time $\tau=2$MTU, averaged over 20 simulations, and over all components of $\xx$. 
The middle panel shows the sensitivity to the observation sampling frequency. For the perfect observation-derived and DA-derived hybrid models, the forecast skill is sensitive to the value of $\Delta t$. The forecast skill is significantly degraded for higher values of $\Delta t$.
This is probably due to the violation of the linear superposition assumption for high values of $\Delta t$ so that the coupling between the resolved part and the unresolved part of the model, as well as the effect of non-linearity of the unresolved part, are no longer negligible.

The right panel of Figure~\ref{fig:luv-p} examines the impact of the observational noise. The result of the reference setting (where the analysis is not filtered before training the NN) is compared with the case of filtering with a rolling mean of 0.05 MTU ({\it i.e.} 5 time steps). Without filtering, the forecast skill deteriorates as the observation noise increases. Filtering the signal for small noise deteriorates the forecast skill too. This means that some source of predictability lies in the fast scales of this model (which confirms results from the middle panel, when it appears that shorter time sampling for observation improves the forecast skill). This small temporal scale variability is damped by the filter, but also by the increase of the observational noise that tends to add randomness on all scales, including the small ones. In this case, except for very strong noise, filtering does not seem to improve drastically the forecast skill. Notably, even a strong noise in the data has only a very small influence on the forecast skill of the hybrid model: when the noise on the observation is multiplied by a factor 20, the error in the forecast at $t_0+2$MTU is only multiplied by a factor 1.3.

From the results on the 2-scales Lorenz model, we conclude that the algorithm is robust against varying data spatial density, but is sensitive to their temporal distribution. 
Also, filtering the analyses obtained from DA may appear useful for slow processes but can deteriorate the results by filtering significant fast processes.

\section{\label{sec:maooam}Application to a low-order coupled ocean-atmosphere model}

\subsection{Description of the model}

We consider here the Modular Arbitrary-Order-Ocean-Atmosphere Model (MAOOAM) introduced by~\cite{decruz2016}. MAOOAM has 3 layers (2 for the atmosphere and 1 for the ocean) and is a reduced-order quasi-geostrophic model resolved in the spectral space. Its state is composed of $n_a$ modes of the atmospheric barotropic streamfunction $\psi_{a,i}$ and the atmospheric temperature anomaly $\theta_{a,i}$ respectively, plus $n_o$ modes of the oceanic streamfunction $\psi_{o,j}$ and the oceanic temperature anomaly $\theta_{o,j}$ respectively. The total number of variables is $N_x = 2n_a + 2n_o$. We consider two versions of MAOOAM:  the true model with dimension $N_z=56$ ($n_a=20$,$n_o=8$) corresponding to the configuration 
``atm. $2x$-$2y$ oc. $2x$-$4y$'' in~\cite{decruz2016} and the truncated model with $N_x=36$ ($n_a=10$,$n_o=8$) corresponding to the configuration  atm. ``$2x$-$4y$ oc. $2x$-$4y$'' in~\cite{decruz2016}.
The truncated model is missing 20 high-order atmospheric variables (10 for the streamfunction and 10 for the temperature anomaly).
Thus the truncated model does not resolve the atmosphere-ocean coupling related to these high order atmospheric modes.


The true model is used to generate synthetic observations. The forecast skill and the long-term properties of the truncated and the hybrid models will be evaluated by inspecting 3 crucial model variables, called \emph{key variables}, that are $\psi_{o,2}$, $\theta_{o,2}$ and $\psi_{a,1}$ ({\it i.e.} the second components of ocean streamfunction and temperature and the first component of the atmospheric streamfunction).
They account for 42\%, 51\%, and 18\% respectively of the variability of a reanalysis of 2-dimensional fields
~\cite{vannitsem2017evidence}, and have been already used in previous studies ({\it e.g.} ~\cite{demaeyer2017stochastic}). MAOOAM has also been recently used to study coupled data assimilation methods~\cite{penny2019strongly,tondeur2020temporal}.
Unsurprisingly, in MAOOAM the ocean variables are considered the slow ones while the atmospheric variables are the fast ones.

\subsection{\label{maooam:setting}Experimental setup}
We will express time in real time units (minutes, hours, days, ...) but, in practice, the model time is non-dimensional. Consequently, the dimensioned time values presented hereafter are not round numbers.

Given the diverse time scales and amplitudes of the MAOOAM variables, the noise parameters are all scaled on a climatological standard deviation of high frequencies $\bsigma^{\rm hf} \in \mathbb{R}^{N_x}$, which is defined as the temporal standard deviation of the state vector after filtering out slow variations of a period longer than 1 month. This high-pass filter is carried out by subtracting the 1-month running average.

The parameters of the experiments are presented in Table~\ref{tab:exp-setting}, in the column labelled MAOOAM.
The true model is integrated over approximately $62$ years after a spinup of $30,000$ years, in the same configuration as in~\cite{decruz2016}.
In all experiments with MAOOAM, the state is fully observed every $27$ hours ($\Delta t=27$ hours) (corresponding to $N_c=1,000$).
A small modification was made to the observations from Eq.\,\eqref{eq:obs} to account for the fact that observations of the ocean are not at the same scale as those of the atmosphere: before being assimilated, instantaneous ocean observations are averaged over a 55 days rolling period centred at the analysis times. 
The EnKF-N is used as DA algorithm. The noise on the model forecast is added only to the atmospheric variables with a standard deviation of $\bsigma^{\rm m}=10^{-3} \bsigma^{\rm hf}_{1:20}$.
As mentioned in section~\ref{sec:method}\ref{method:algo}, the analysis obtained from the DA is filtered. The slow processes are expected to occur mainly in the ocean, so only the ocean components of the state vector $\xx^{\rm a}_{0:K}$ are filtered to produce $\xx^{\rm s}_{0:K}$. Differently from the L2S model experiments, filtering the analysis has proven necessary to train the hybrid model using MAOOAM.

The NN-architecture is a simple 3 layers multi-layer perceptrons; see Table~\ref{tab:exp-setting} for full details. As opposed to the L2S model, the state vector has no locality properties (because it is defined in the spectral space), so the convolutional layers are not applicable (see the discussion about locality in~\cite{brajard2020combining}). The training of the NN is performed in the same way as for the L2S experiments.

\subsection{Results}
The forecast skill metrics are presented in Table~\ref{tab:maooam-forecast} for the truncated model as well as for the perfect observation-derived and the DA-derived hybrid models. Given the different time scales involved, the forecast lead time of the key atmospheric variable $\psi_{a,1}$ is 1 day whereas the forecast lead time of the two key oceanic variables $\psi_{o,2}$ and $\theta_{o,2}$ is 2 years. It can be seen that both perfect observation-derived and DA-derived hybrid models have superior skill to the truncated model. The improvement is larger for the ocean, with a factor of 2 to 3, and is similar for both hybrid models.
Recall that the true model has here the same oceanic variables as the truncated model, so there is no difference in the representation of the pure oceanic processes.
The improvement is thus fully due to an enhanced representation of the atmosphere-ocean coupling processes, the hybrid model better representing the interplay between the unresolved fast atmospheric variables and the slow oceanic variables.

The atmospheric key variable is improved to a lesser extent by the two hybrid models, and the perfect observations-derived model is significantly better than the DA derived model. This proves the limited capability of the hybrid model to represent a fast process, a situation further exacerbated in the case of the DA-derived hybrid model, when only noisy and partial data are at disposal. This result was indeed expected given the assumptions made on the unresolved term of the model in section~\ref{sec:method}\ref{method:approx} when a slow variation of the unresolved term was assumed. The fast processes are also less accurately represented because the sampling rate of the observations ($27$ hours) is well beyond the atmospheric time scale, and because of the presence of the observation errors (when applied).

\begin{figure}[!ht]
\centering
\includegraphics[width=.9\textwidth]{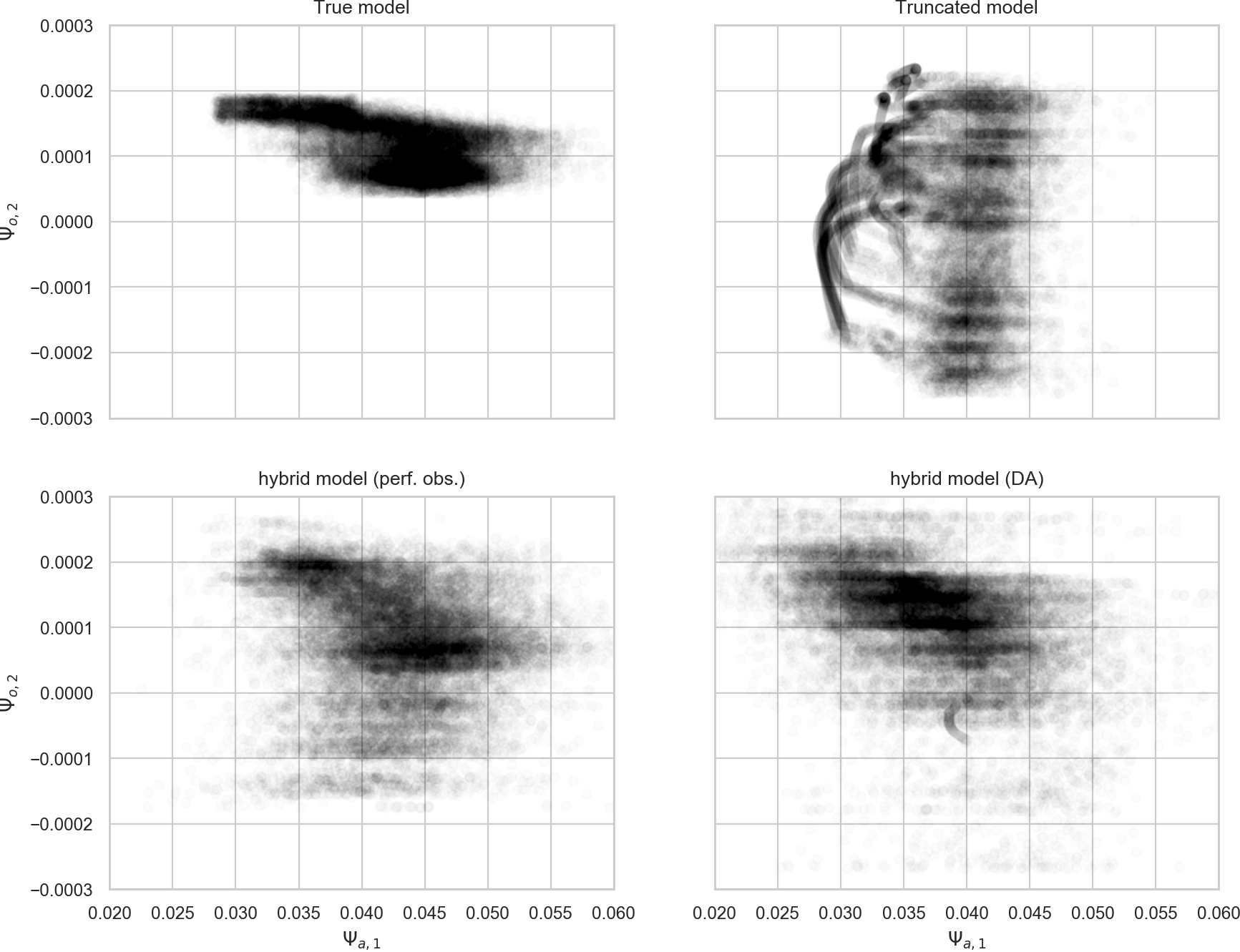}
\caption{Cross-section of the attractor for two key variables $\psi_{a,1}$ and $\psi_{o,2}$ in the true model (upper left), the truncated model (upper right), the perfect-observation-derived hybrid model (lower left) and the DA-derived hybrid model (lower right).}
\label{fig:maooam-attractor}
\end{figure}

In Figure~\ref{fig:maooam-attractor}, the attractors of the different models are displayed in the phase space defined by two key variable: $\psi_{o,2}$ and $\psi_{a,1}$. A significant difference can immediately be seen between the attractors of the truncated and the true models: the truncated model visits areas of the phase space that are not admitted in the real dynamics. Remarkably, these discrepancies are reduced by the hybrid models both derived from perfect observation and from DA. Some states seem to remain out of the true model attractor, however, but much fewer.

Quantitative characterisation of the attractors is presented in Table~\ref{tab:maooam-stats}, which provides the 3 quartiles (including the median) for each key variable. For the truncated model and the hybrid models, the difference of the quartiles is given relatively to the true model. For all the oceanic variables the distribution of the values of both hybrid models is significantly closer to the true distribution than for the truncated model. For the key atmospheric variable $\psi_{a,1}$, only the hybrid model derived from perfect observations shows an improvement. It confirms the conclusion made on the forecast skill that the hybrid model represents well the slow process, in particular oceanic variables in this case, and that the fast processes are not fully retrieved, in particular in case of the DA-derived hybrid model.

\begin{table}[!ht]
\caption{Forecast R-RMSE of hybrid and truncated MAOOAM models}
\label{tab:maooam-forecast}
\begin{tabular}{|c|ccc|}
\hline
R-RMSE(lead time $\tau$) & $\psi_{o,2}$(2 years) & $\theta_{o,2}$(2 years) & $\psi_{a,1}$(1 day)\\
Truncated & 0.23 & 0.21 & 0.36 \\
Perfect obs. hybrid & 0.07 & 0.07 & 0.23 \\
DA hybrid & 0.10 & 0.06 & 0.28 \\
\hline
\end{tabular}
\vspace*{-4pt}
\end{table}

\begin{table}[!ht]
\caption{Quartiles of the key variables for the MAOOAM model relative to the true model.}
\label{tab:maooam-stats}
\begin{tabular}{c|ccc|ccc}
\cline{2-7}

&\multicolumn{3}{c|}{$\psi_{o,2}$} &\multicolumn{3}{c}{$\theta_{o,2}$} \\
\cline{2-7}
& Q1 & M & Q3 & Q1 & M & Q3 \\
\hline
True model &
$7.8\cdot10^{-5}$ & $1.1\cdot10^{-4}$ & $1.5\cdot10^{-4}$ &
$8.2\cdot10^{-2}$ & $1.2\cdot10^{-1}$ & $1.4\cdot10^{-1}$ \\
Truncated &
-229\% & -80\% & -26\% &
-22\% & -10\% & -6\% \\
Perfect obs. hybrid & -55\% & -26\% & 0.6\% & 7\% & -2\% & -4\% \\
DA hybrid & -14\% & 9\% & 8\% & 8\% & -5\% & -0.2\% \\
\hline
&\multicolumn{3}{c|}{$\psi_{a,1}$} \\
\cline{2-4}
& Q1 & M & Q3 \\
\cline{1-4}
True model & $3.9\cdot10^{-2}$ & $4.3\cdot10^{-2}$ & $4.6\cdot10^{-2}$\\
Truncated & -12\% & -11\% & -11\%\\
Perfect obs. hybrid & -0.6\% & -2\% & 0.02\%\\
DA hybrid & -15\% & -14\% & -11\% \\
\cline{1-4}
\end{tabular}%
\vspace*{-4pt}
\end{table}

\section{Conclusion}
We have developed a novel method to build a hybrid model consisting of a physics-based truncated model and a data-driven model of the unresolved processes. The approach is based on realistic assumptions that only noisy and incomplete observations are available at a lower frequency than the model integration time step. 

With a two-scale low-order chaotic system \cite{lorenz2005designing}, we showed that the hybrid model forecast skill is sensitive to the observation frequency but very robust against high observational noise and sparse spatial distribution. This is probably due to the fact that reduced observation frequencies challenge the validity of the linear superposition assumption more than large observational noise (see the discussion in Appendix of~\cite{bocquet2017four}).  
We then applied the method to the low-order coupled ocean-atmosphere model MAOOAM~\cite{decruz2016} which contains multiple temporal scales. Forecast skill and global statistics were significantly improved by the hybrid model compared with the truncated model encouraging further studies to high-dimensional and more realistic scenarios. Notably, the hybrid model derived from noisy observations has comparable forecast skill on the oceanic variables to that of the hybrid model derived from perfect observations.

In view of operational systems, it should be noted that the proposed algorithm relies on two existing data assimilation and neural networks training techniques that both scale well in high-dimension (see, {\it e.g.},~\cite{sakov2012topaz4} and~\cite{lecun2015deep}). In principle, the present algorithm can be applied to larger and more realistic problems. In particular, the fact that the method does not rely on the adjoint of the truncated model is an advantage in terms of code maintenance. However, we foresee some practical challenges: for instance, the computational architecture and the data types used for physics-based numerical models and for machine learning algorithms can be very different ({\it e.g.} multi-core supercomputers and graphics processing units). Training and running hybrid models efficiently imposes heavy requirements on both the hardware and software and may come with an overhead even if some tools are very promising~\cite{ott2020fortran}. 

The approach presented here can also accommodate the additional representation of the remaining model error ({\it i.e.} the model error of the hybrid model):  it could either be done within the numerical model by parameterising the model error~\cite{bocquet2020bayesian}, or by training stochastic neural networks~\cite{gal2016dropout}.

\vskip6pt

\enlargethispage{20pt}


\dataccess{Data used in all the experiments are available at \url{ftp://ftp.nersc.no/reddaml/}. The instructions to download the data, to run all the Lorenz 2 scale application and to reproduce figures and tables of the article can be found at \url{https://github.com/brajard/reddaml/releases/tag/v1.0}.}

\aucontribute{JB first proposed the theory, implemented and conducted the numerical experiments. All authors have contributed to
the interpretation of the theory and the results as well as the edition of the manuscript. All authors approved the manuscript.}

\competing{The authors declare that they have no competing interests.}

\funding{JB and LB have been funded by the project REDDA (\#250711) of the Norwegian Research Council. JB has also been funded by the project SFE (\#2700733) of the Norwegian Research Council. AC has been funded by the UK Natural Environment Research Council award NCEO02004. CEREA is member of Institut Pierre–Simon Laplace (IPSL).}

\ack{The authors are thankful to Patrick Raanes (NORCE, NO) for his support on the use of the data assimilation python platform DAPPER, and to Jonathan Demaeyer (RMI, BE) for the insightful discussions about MAOOAM. Thanks to Jos\'ephine Schmutz (IMT) for his review of the GitHub repository. Many thanks also to the two anonymous reviewers for their valuable comments.}



\ifdefined\isbibtex
\bibliographystyle{rsta}
\bibliography{references}
\else

\fi

\end{document}